\documentclass[aps]{revtex4-1}
\usepackage{graphicx}
\usepackage{amsmath}
\usepackage{amssymb}
\usepackage{tikz}
\usepackage{float}
\usepackage{sidecap}
\usepackage[caption = false]{subfig}
\usepackage{tabularx,booktabs}
  \newcolumntype{Y}{>{\centering\arraybackslash}X}
\usepackage[margin=1in]{geometry}
\usetikzlibrary{snakes}
\usepackage[utf8]{inputenc}
\usepackage[english]{babel}
\usepackage{multirow}

\begin{document}
\preprint{APS/123-QED}
\title{Study of the long-lived excited state in the neutron deficient nuclides $^{195,197,199}$Po by precision mass measurements}
\author{N.~A.~Althubiti}
\affiliation{School of Physics and Astronomy, The University of Manchester, Manchester, United Kingdom}
\affiliation{Al-jouf University, Sakakah, Saudi Arabia;}
\author{D.~Atanasov}
\affiliation{Max-Planck-Institut für Kernphysik, Heidelberg, Germany;} 
\author{K.~Blaum}
\affiliation{Max-Planck-Institut für Kernphysik, Heidelberg, Germany;} 
\author{T.~E.~Cocolios}
\affiliation{School of Physics and Astronomy, The University of Manchester, Manchester, United Kingdom}
\affiliation{KU Leuven, Instituut voor Kern- en Stralingsfysica, Leuven, Belgium;}
\author{T.~Day Goodacre}
\affiliation{School of Physics and Astronomy, The University of Manchester, Manchester, United Kingdom,}
\affiliation{EN Department, CERN, Geneva, Switzerland} 
\affiliation {TRIUMF, Vancouver (BC), Canada;}
\author{G.~J.~Farooq-Smith}
\affiliation{School of Physics and Astronomy, The University of Manchester, Manchester, United Kingdom}
\affiliation{KU Leuven, Instituut voor Kern- en Stralingsfysica, Leuven, Belgium;}
\author{D.~V.~Fedorov}
\affiliation{Petersburg Nuclear Physics Institute, NRC Kurchatov Institute, Gatchina, Russia; }
\author{V.~N.~Fedosseev}
\affiliation{EN Department, CERN, Geneva, Switzerland;}
\author{S.~George}
\affiliation{Max-Planck-Institut für Kernphysik, Heidelberg, Germany;} 
\author{F.~Herfurth}
\affiliation{GSI Helmholtzzentrum für Schwerionenforschung GmbH, Darmstadt, Germany;} 
\author{K.~Heyde}
\affiliation{Department of Physics and Astronomy, Ghent University, Ghent, Belgium;} 
\author{S.~Kreim}
\affiliation{Max-Planck-Institut für Kernphysik, Heidelberg, Germany} 
\affiliation{EP Department, CERN, Geneva, Switzerland;}
\author{D.~Lunney}
\affiliation{CSNSM-IN2P3-CNRS, Orsay, France;}
\author{K.~M.~Lynch}
\affiliation{EP Department, CERN, Geneva, Switzerland;}
\author{V.~Manea}
\affiliation{EP Department, CERN, Geneva, Switzerland;}
\author{B.~A.~Marsh}
\affiliation{EN Department, CERN, Geneva, Switzerland;} 
\author{D.~Neidherr}
\affiliation{GSI Helmholtzzentrum für Schwerionenforschung GmbH, Darmstadt, Germany;} 
\author{M.~Rosenbusch}
\affiliation{Ernst-Moritz-Arndt-Universität, Institut für Physik, Greifswald, Germany;}
\author{R.~E.~Rossel}
\affiliation{EN Department, CERN, Geneva, Switzerland} 
\affiliation{Institut für Physik, Johannes Gutenberg Universität, Mainz, Germany;}
\author{S.~Rothe}
\affiliation{EN Department, CERN, Geneva, Switzerland;} 
\author{L.~Schweikhard}
\affiliation{Ernst-Moritz-Arndt-Universität, Institut für Physik, Greifswald, Germany;}
\author{M.~D.~Seliverstov}
\affiliation{EN Department, CERN, Geneva, Switzerland}
\affiliation{Petersburg Nuclear Physics Institute, NRC Kurchatov Institute, Gatchina, Russia; }
\author{A.~Welker}
\affiliation{EP Department, CERN, Geneva, Switzerland}
\affiliation{Institut für Kern- und Teilchenphysik, Technische Universität Dresden, Germany;} 
\author{F.~Wienholtz}
\affiliation{Ernst-Moritz-Arndt-Universität, Institut für Physik, Greifswald, Germany;}
\author{R.~N.~Wolf}
\affiliation{Max-Planck-Institut für Kernphysik, Heidelberg, Germany} 
\affiliation{ARC Centre of Excellence for Engineered Quantum Systems, School of Physics, The University of Sydney, Australia.} 
\author{K.~Zuber}
\affiliation{Institut für Kern- und Teilchenphysik, Technische Universität Dresden, Germany;}

\date{\today}

\begin{abstract}
Direct mass measurements of the low-spin $3/2^{(-)}$ and high-spin $13/2^{(+)}$ states in the neutron-deficient isotopes $^{195}$Po, $^{197}$Po, and high-spin $13/2^{(+)}$ state in $^{199}$Po were performed with the Penning-trap mass spectrometer ISOLTRAP at ISOLDE-CERN.  These measurements allow the determination of the excitation energy of the isomeric state arising from  the $\nu$i$_{13/2}$ orbital in $^{195,197}$Po.  Additionally, the excitation energy of isomeric states of lead, radon, and radium isotopes in this region were obtained from $\alpha$-decay chains. The new excitation energies complete the knowledge of the energy systematics in the region and confirm for the first time that the $13/2^{(+)}$ states remain isomeric, independent of the number of valence neutrons.
\end{abstract}

\maketitle

\section{\label{sec:level1}Introduction}
In the neutron-deficient lead region,  nuclei around the proton number  $Z=82$ and below  the neutron number $N=126$ are known to exhibit shape coexistence \cite{1}. A striking example is $^{186}$Pb \cite{2}, where spherical, oblate and prolate states were found to compete all within  700 keV energy difference. Shape coexistence arises from the possibility that proton-pair excitations across the $Z=82$ proton closed shell can be strongly lowered in their excitation energy when moving away from the neutron-closed shell at $N=126$. Pairing plays a particularly important role in lifting protons from the low-j 3s$_{1/2}$ and 2d$_{3/2}$ orbitals, below $Z=82$, into the high-j 1h$_{9/2}$ orbital above $Z=82$. The very strong deformation-driving proton-neutron interaction between these proton-pairs and the increasing number of valence holes (in particular for the Pb region) has resulted in strong evidence of collective bands with rotational characteristics, with a minimal excitation energy for the band head near mid shell at $N=104$ \cite{Julin2001,Elseviers2011,Rahkila2010,Kesteloot2015,VandeVel2003}. 

\begin{figure}[t!]
\centering
\includegraphics[width=0.5\textwidth]{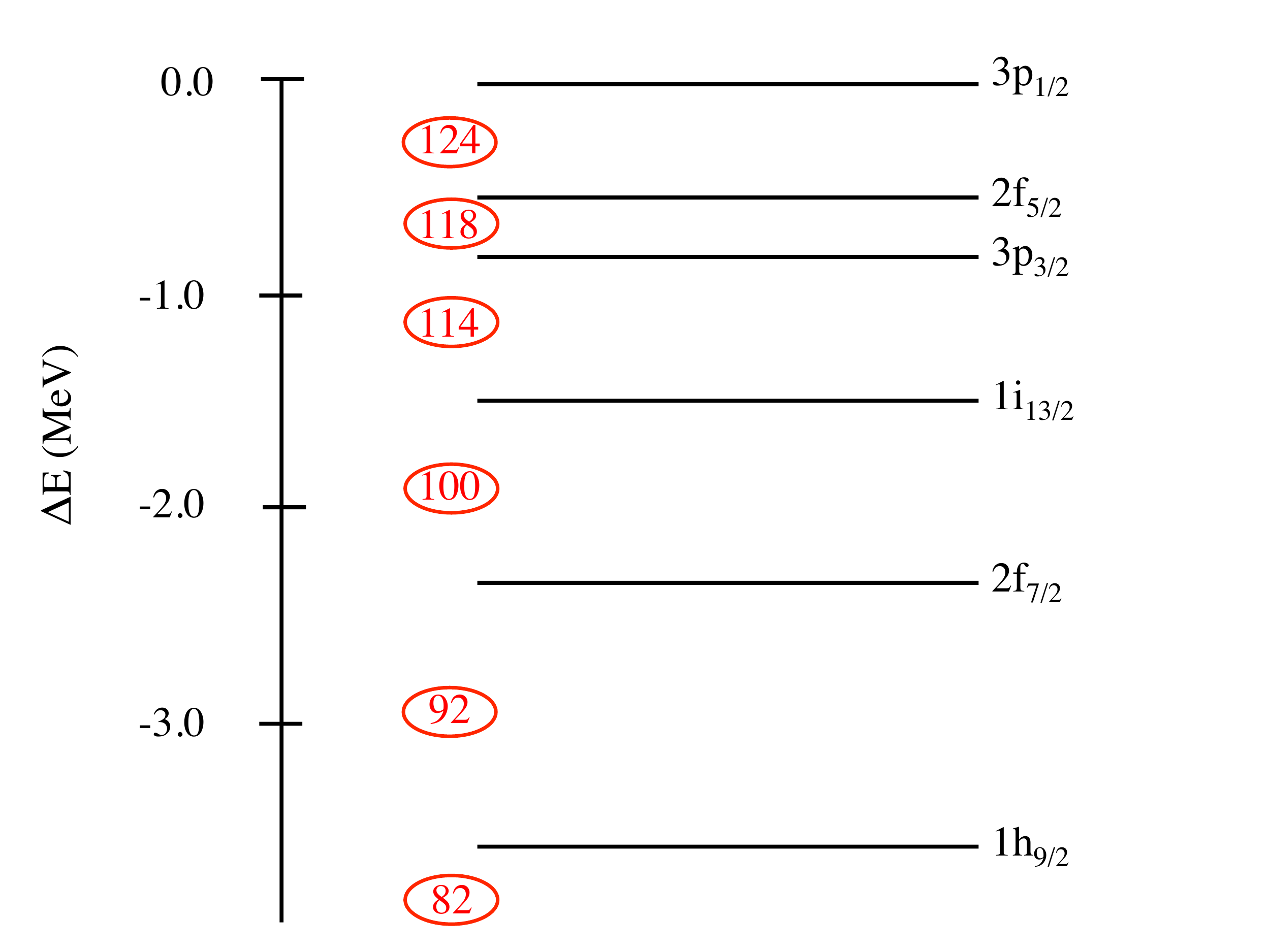}
\caption{\label{fig:1p} (color online) The neutron spherical shell model orbitals, their energies taken from  experimental data on $^{207}$Pb \cite{NDS207}, as well as the neutron number corresponding to the filled orbital(s).}
\end{figure}

When studying the expected low-lying states for the odd-mass ($Z=84$) Po nuclei and moving away from the $N=126$ closed shell
towards mid-shell at $N=104$, within a simple shell-model approach, one first encounters the low-spin 3p$_{1/2}$, 2f$_{5/2}$ and 3p$_{3/2}$ orbitals before the high-spin 1i$_{13/2}$ orbital, with opposite parity (see Fig. \ref{fig:1p}). In this simple approach, one would expect that sequentially increasing the number of holes within the $N=126$ closed shell, the ground-state spin of even-$Z$, odd-$N$ isotopes would change over the 1/2$^-$, 5/2$^-$, 3/2$^-$ sequence, reaching a spin of 13/2$^+$ for the ground-state for $N<114$. The neutron orbits of the shell model in this region are shown in Fig. \ref{fig:1p}. A  similar situation occurs in the case of the odd-mass Sn nuclei, where the spherical shell model would suggest that from $A=Z+N=121$ to $A=132$, in the vicinity of the $Z=50$ shell closure, the high-spin 1i$_{11/2}$ orbital gives rise to a 11/2$^-$ ground state. Such a situation is, however, only observed for the $A=123$ to $A=127$ nuclei (i.e. $N=73$ to $N=77$) \cite{Anselment1986,LeBlanc2005}. 

It is clear that to have a better description of the changing low-spin excited states in the odd-mass Po nuclei, the pairing properties in different orbitals will play an essential role. Pairing in a given $J^\pi=0^+$ state resulting from two nucleons with an identical angular momentum ($j_1=j_2=j$) is:
\begin{equation}
\label{M0}
\Delta E_{pairing} (j^2;0^+)=-G\cdot(j+1/2),
\end{equation}
where $G$ is the coupling strength \cite{p.ring}. This energy gain therefore varies linearly with increasing $j$. This property will, when removing neutrons from the $N=126$ core, try to form pairs within the 1i$_{13/2}$ orbital, and thus, in the region with $N\sim116$ overturn the standard ordering for most of the ground states. Experimental data in the odd-$A$ nuclei demonstrate that the high-spin 13/2$^+$ state is still an excited state before becoming almost degenerate with the low-spin 3/2$^-$ state halfway between the shell-closures $N=82$ and $N=126$ \cite{3,4}. 
A blind shell-model approach would go as follows: in going from $N=115$ to $N=113$, it would be expected that the last neutron in the 3p$_{3/2}$ orbital and a neutron from the 1i$_{13/2}$ orbital are removed, at the expense of breaking a pair in that orbital, thereby forming a 13/2$^+$ ground state. If, however, the valence neutron is kept in the 3p$_{3/2}$ orbital while removing two neutrons from 1i$_{13/2}$ orbital, this results in a $|${3p$_{3/2}$ (1i$_{13/2}) ^{-2}_{0^+}$ ;$3/2^-$} $\rangle $ configuration.

In case the single-particle energy difference between these two orbitals is smaller than the pairing gain in the 1i$_{13/2}$ orbital, the 13/2$^+$ state can remain an excited state over a long chain of isotopes.

Similarly for odd-odd isotopes, a competition between the neutron $13/2^{(+)}$ state and $3/2^{(-)}$ state was observed in even-$A$ thallium ($Z=81$),  bismuth ($Z=83$) and actinium ($Z=85$) isotopes \cite{5,6} where the  level spacing between states decreases as $N$ is reduced. The isomeric states $7^{+}$ ($10^{-}$)  that originate from $\pi s_{1/2} \otimes \nu i_{13/2}$  ($\pi h_{9/2} \otimes \nu i_{13/2}$) configurations, were found to be above the ground states $2^{-}$ ($3^{+}$), that arise from $\pi s_{1/2} \otimes \nu p_{3/2}$ ($\pi h_{9/2} \otimes \nu p_{3/2}$) configurations, by a few tens of keV.

The $13/2^{(+)}$ state has been tracked in previous studies by means of delayed electron and $\gamma$ spectroscopy, however the information about the level spacing between the $13/2^{(+)}$ and $3/2^{(-)}$ states is not complete near $N=104$, especially in odd-$A$ lead and polonium isotopes where the states are probably almost degenerate. Most importantly, it is not known whether the 13/2$^{(+)}$ state becomes the ground state near $N=104$ as a naive interpretation of the spherical shell model would indicate. This information will give insight into the specific neutron single-particle energies, as well as into the interplay of the pairing energies in the different orbitals and the proton-neutron interaction energy. As observed in the odd-mass Sn nuclei, one expects that eventually the 13/2$^{(+)}$ state will become the ground state for a still lower neutron number beyond mid-shell. This is, however, not within reach experimentally. A full pairing study of the neutron system could, on the other hand, extract the corresponding one quasi-particle energies. 

The relation with known data on nuclear charge radii can shed light on the onset of various nuclear shapes that are known to appear near the neutron mid-shell region such as the competition between oblate and prolate deformation. Hartree-Fock-(Bogoliubov) calculations, also going beyond the mean-field description, have shown that the even-even Po nuclei exhibit an oblate $0^+$ ground state down to $N=108$ ($A=192$), changing towards the prolate shape from $N=106$ ($A=190$) and lower \cite{Yao,Bender,Rodri,Grahn08,Grahn09}.  These results are in line with studies of the energy surfaces and particle-plus-rotor (PPR) calculations \cite{Andreyev06,4,Vanvel} starting from a more phenomenological deformed Woods-Saxon potential \cite{Satula}. The $\gamma$-ray energy spectra in the $^{193,195,197}$Po nuclei show a pronounced rotational-aligned band on top of the 1i$_{13/2}$ state, pointing towards effects that may be best understood if a slight oblate shape is the one determining the lowest-lying states \cite{Fotiades,4}. This is the region where low $\Omega$ (the projection of the total angular momentum along the symmetry axis) values originating from the 1i$_{13/2}$ are near the neutron Fermi level at $N=109,111,113$ with an increasing projection with decreasing neutron number. It is only when reaching the $^{190}$Po nucleus, and even lower neutron numbers (and $A$), that most probably, the nuclear ground-state deformation changes towards prolate and the subsequent strongly-coupled rotational bands become the fingerprint to signal this transition. The data on charge radii \cite{Kesteloot2015,Cocolios2011,Seliverstov2013} are consistently pointing towards an increase relative to the droplet model which is starting at $N = 114$ ($A=198$) and moving quickly up towards the mid-shell point near $N=104$ ($A=188$) where a prolate ground-state is expected \cite{Yao}.

In this respect, the measurement of the excitation energy of the level arising from  $\nu i_{13/2}$ orbital is of interest to bridge the knowledge gap and to track the evolution of nuclear structure in this region. Since no decay path exists between these states, the measurement of the excitation energy demands precise mass values of the isomeric and ground states. Thus mass measurements of the isotopes of interest were performed with the Penning-trap mass spectrometer at ISOLTRAP, reaching relative uncertainties of the order of $10^{-8}$ \cite{8,2013S}. The identification of the investigated state was achieved by utilizing the resonant ionization laser ion source (RILIS) \cite{9} in combination with $\alpha$-decay-spectroscopy measurements.

In this article, we report on the high-resolution mass measurement of neutron-deficient polonium isotopes with the ISOLTRAP mass spectrometer at CERN-ISOLDE. Isomer-selective laser ionisation and decay spectroscopy were combined with ISOLTRAP's mass measurements capabilities to reveal the state ordering and excitation energy of odd-$A$ isotopes. Through mass links determined by $\alpha$-decay chains, the state ordering and excitation energies in odd-$A$ lead, radon, radium are also determined.
\begin{figure}[t!]
\centering
\includegraphics[width=0.8\textwidth]{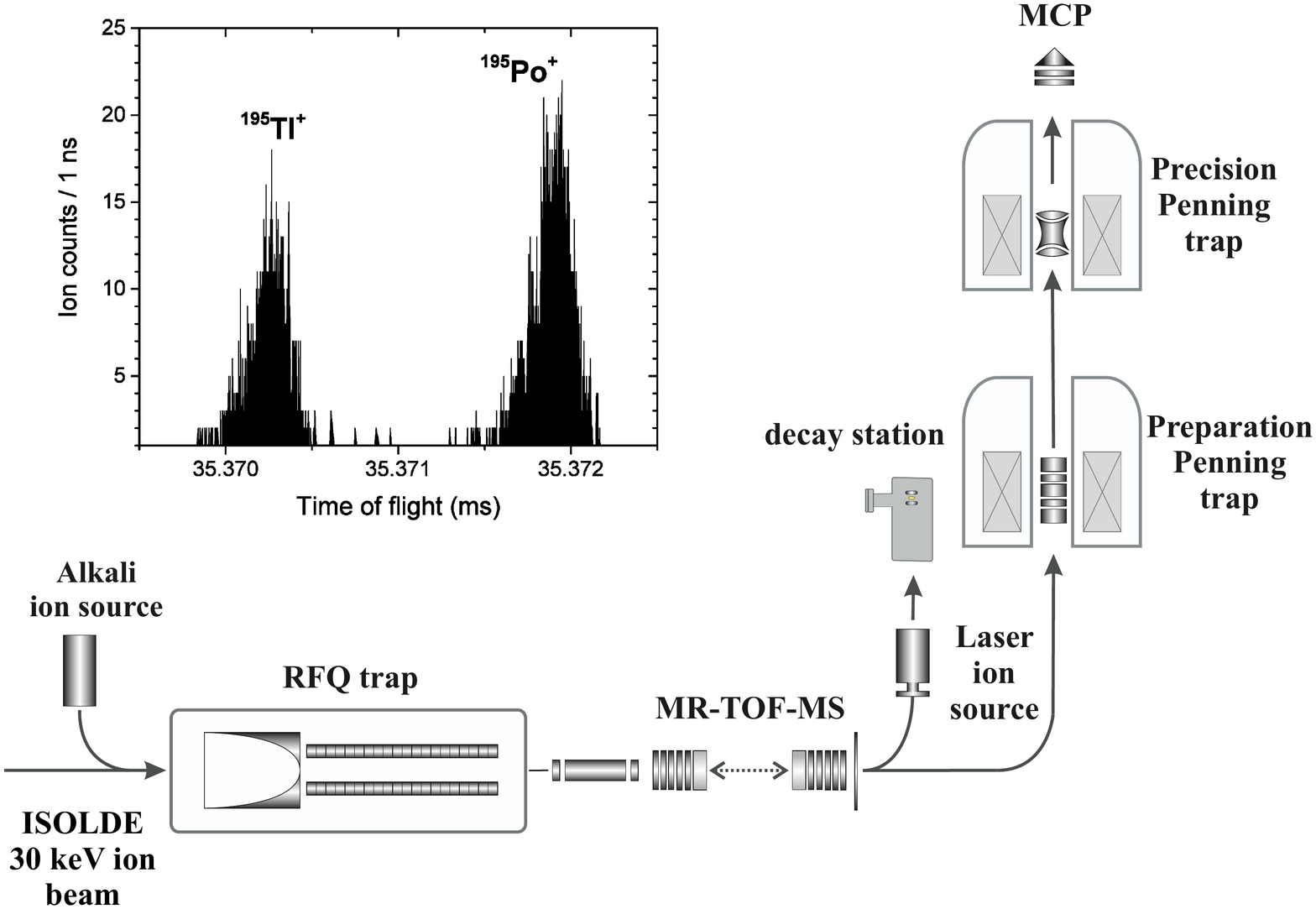}
\caption{\label{fig:1} Schematic view of the ISOLTRAP setup with the $\alpha$-decay station. Inset: time-of-flight spectrum of mass A=195. Full separation of $^{195}$Po$^+$ from $^{195}$Tl$^+$ isotopes was achieved after 1000 revolutions in the MR-ToF MS.}
\end{figure}
\section{\label{ExSet}Polonium production and Experimental procedure}
The neutron-deficient polonium isotopes were produced at the online isotope separator ISOLDE-CERN \cite{10}. A thick uranium carbide (UC$_x$) target was bombarded with 1.4 GeV protons delivered from the CERN Proton Synchrotron Booster (PSB). Proton-induced spallation reactions in the UC$_x$ target led to the production of the different polonium isotopes.  The radioactive recoils diffused from the target material and effused via a hot cavity (2000 $^{\circ}$C) toward the Resonant Ionization Laser Ion Source (RILIS) \cite{9}, where they were ionized. The high temperature of the cavity also resulted in the ionization of isobaric contamination with low ionisation potentials, such as thallium, francium and bismuth. At mass A=197 an additional contamination from $^{181}$Ta$^{16}$O  was present, which was related to the use of the high-power non-resonant 532~nm laser. An alternative ionization scheme with only resonant transitions was investigated. No autoionizing states were found when scanning the range 544-588~nm. Instead, a scan for Rydberg states was performed, and finally the optimal scheme that was used was 256~nm $\rightarrow$ 843~nm$\rightarrow$ 594~nm$^{(Rydberg)}$, to an atomic level at an energy of 67772~cm$^{-1}$. The efficiency of this scheme was up to a factor of 10 lower than that of the 256~nm$\rightarrow$ 843~nm $\rightarrow$ 532~nm$^{(non-resonant)}$ scheme. Nonetheless, the reduction in the $^{181}$Ta$^{16}$O was sufficient to enable the measurements at mass A=197 to take place at ISOLTRAP.

By using a narrow-band laser frequency for the 843~nm transtion \cite{Rot13_NB}, the enhancement of one isomer over the other was possible in odd-$A$ isotopes \cite{11,Cocolios2011Po199}. However, the complex hyperfine structure of these polonium isotopes prevents the production of completely isomerically pure beams \cite{Seliverstov2013}. The polonium isotopes were accelerated up to 30 keV and directed through the high-resolution separator HRS dipole magnets, separated according to the mass-to-charge ratio, and deflected to ISOLTRAP 
where the decay and mass measurements were performed.

Figure \ref{fig:1} shows a schematic view of the ISOLTRAP setup. 
It  consists of four traps each of which has a specific function. Usually the experimental cycle at ISOLTRAP contains three processes: cooling and bunching, isobaric purification and finally, frequency measurements. Additionally, decay spectroscopy was performed in order to distinguish between the ground and isomeric states and to determine the yield ratio. The measurements were thus divided into two parts: decay spectroscopy and mass spectrometry. The ion beam from ISOLDE was captured, cooled and bunched in a gas filled radio-frequency-quadrupole (RFQ) trap \cite{14}. The resulting ion bunches were then ejected to ISOLTRAP's multi-reflection time-of-flight mass spectrometer (MR-ToF MS)\cite{15}.

In the mass region of interest, with $A=195,197,199$, thallium isotopes are the main isobaric contaminants. The resolving power needed to separate polonium from the thallium isobars is $\approx 10^4$, which is within the resolving power of the MR-ToF MS (m/$\Delta m=10^5$ after 30\,ms) (see inset in Fig. \ref{fig:1}) \cite{Wienholtz}. The pure polonium samples were then ejected from the MR-ToF MS to either the $\alpha$-decay-spectroscopy station (DSS2.0) \cite{13} or to the Penning traps. An electrostatic quadrupole bender was used to deflect the ion bunches by $90^{\circ}$ to the DSS2.0, where the polonium ions were implanted into a $20\,\mu $g\,cm$^{-2}$ carbon foil fixed to a manual linear actuator. An MCP detector was also attached to the linear actuator for beam tuning. The carbon foil was surrounded by two silicon detectors to identify the energy of the charged particles emitted from the decay of the polonium isotopes. The solid angle coverage of the silicon detectors was $64\%$. The events were recorded with a triggerless data acquisition system providing the energy and timestamp of each event. Additional timing information was provided from the impact of the protons on the ISOLDE target and release of an ion bunch from the MR-ToF MS.

\begin{figure*}[t!]
\centering
\begin{tabular}{cc}
\subfloat[ ]{\includegraphics[width = 0.5\textwidth]{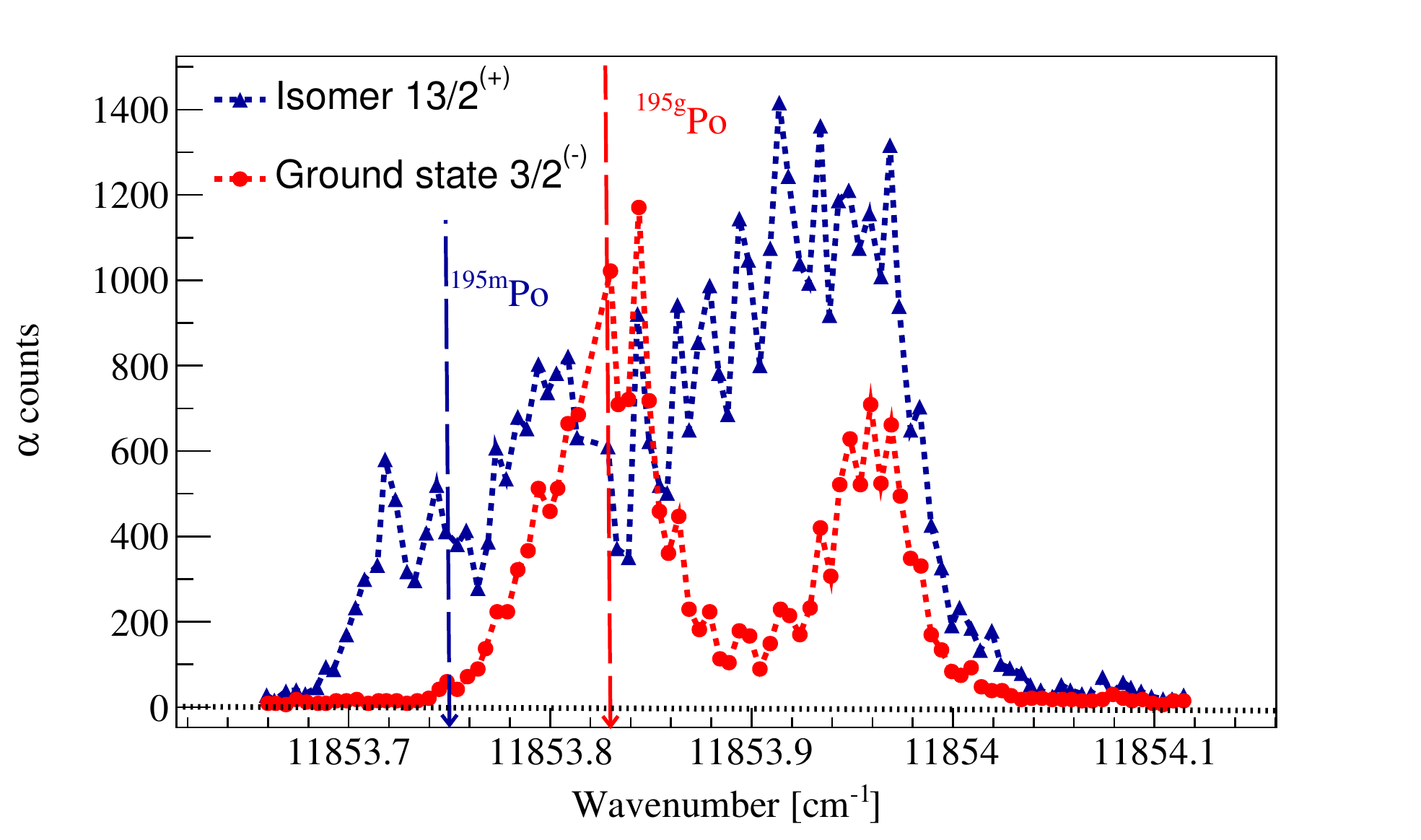}}
\subfloat[ ]{\includegraphics[width = 0.5\textwidth]{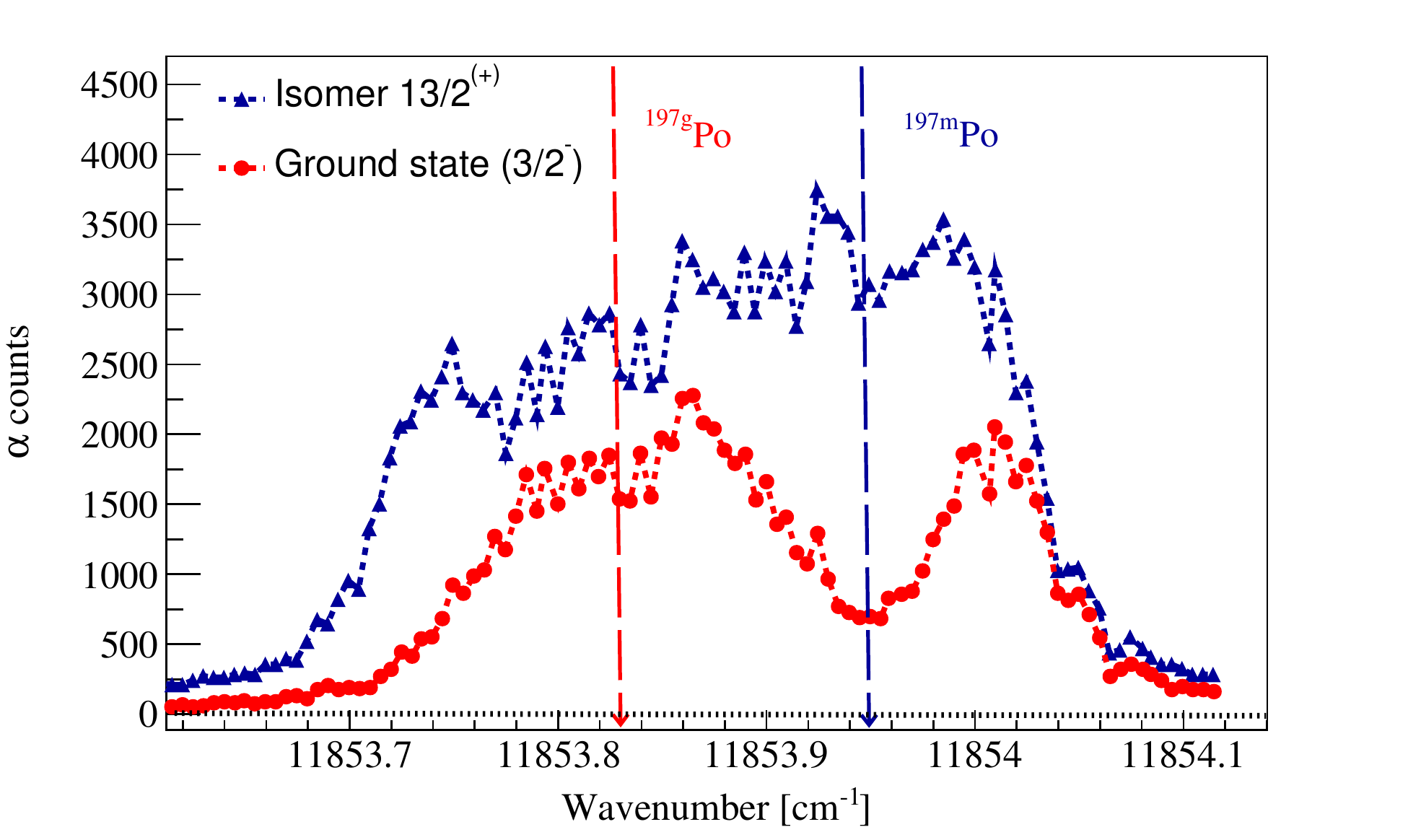}}
\end{tabular}
\caption{\label{fig:2} (color online) The hyperfine structure of the 3/2$^{(-)}$ and 13/2$^{(+)}$ states in $^{195}$Po (left) and $^{197}$Po (right). The dashed lines indicate the wavenumbers used for the mass measurements \cite{Seliverstov2013,11}. Note that the two hyperfine spectra were recorded in a different experiment, hence the laser resolution achieved in the present work  is different.}
\end{figure*}

Alternatively, the ions could be delivered from the MR-ToF MS to the preparation Penning trap for further purification where the mass-selective resonant buffer-gas cooling technique was applied \cite{16}. The damping of the unwanted ion motions (axial and radial motions) in the preparation Penning trap provides an optimal injection of the ions into the precision Penning trap where the mass determination took place.
In the precision trap, the mass was  measured by using the time-of-flight ion-cyclotron-resonance (ToF-ICR) technique \cite{17}. A quadrupolar  RF field was scanned around the cyclotron frequency of the ions under investigation \cite{18}. For each frequency, the ions were then ejected towards an MCP detector and their time-of-flight measured. 
When the frequency of the RF field coincide with the cyclotron frequency of the ion of interest, the time of flight reaches a minimum. By measuring the cyclotron frequency of the ion, one can extract its charge-to-mass ratio mass $q/m$ from its cyclotron frequency:
\begin{equation}
\nu_c = \frac{qB}{2\pi m}.
\label{M0}
\end{equation}

To eliminate the magnetic field $B$ from the equation (as it cannot be directly measured with sufficient accuracy), cyclotron frequency measurements $\nu_{c,ref}$ of a reference ion with a precisely known atomic mass $m_{ref}$ were performed. The atomic mass $m_{x}$ of the nuclide of interest is then obtained using the cyclotron-frequency ratio, $r={\nu_{c,ref}}/{\nu_c}$, as:
\begin{equation}
\label{M1}
m_{x}= {r(m_{ref}-m_e)+m_e}.
\end{equation}

\section{\label{Res}Results}

During the polonium campaign, two different types of measurements were performed. First, the polonium isotopes were ionized with RILIS and sent to the decay station where the accurate ratio of ground state to isomer was obtained from $\alpha$-decay schemes. Second, for mass measurements, ToF-ICR spectra were recorded for each individual state or for both together. The combination of these different measurements enabled determination of the precise mass values of the investigated states.

\subsection{$^{195}$Po}

\begin{figure*}
\centering
\begin{tabular}{cc}
\subfloat[ ]{\includegraphics[width = 0.45\textwidth]{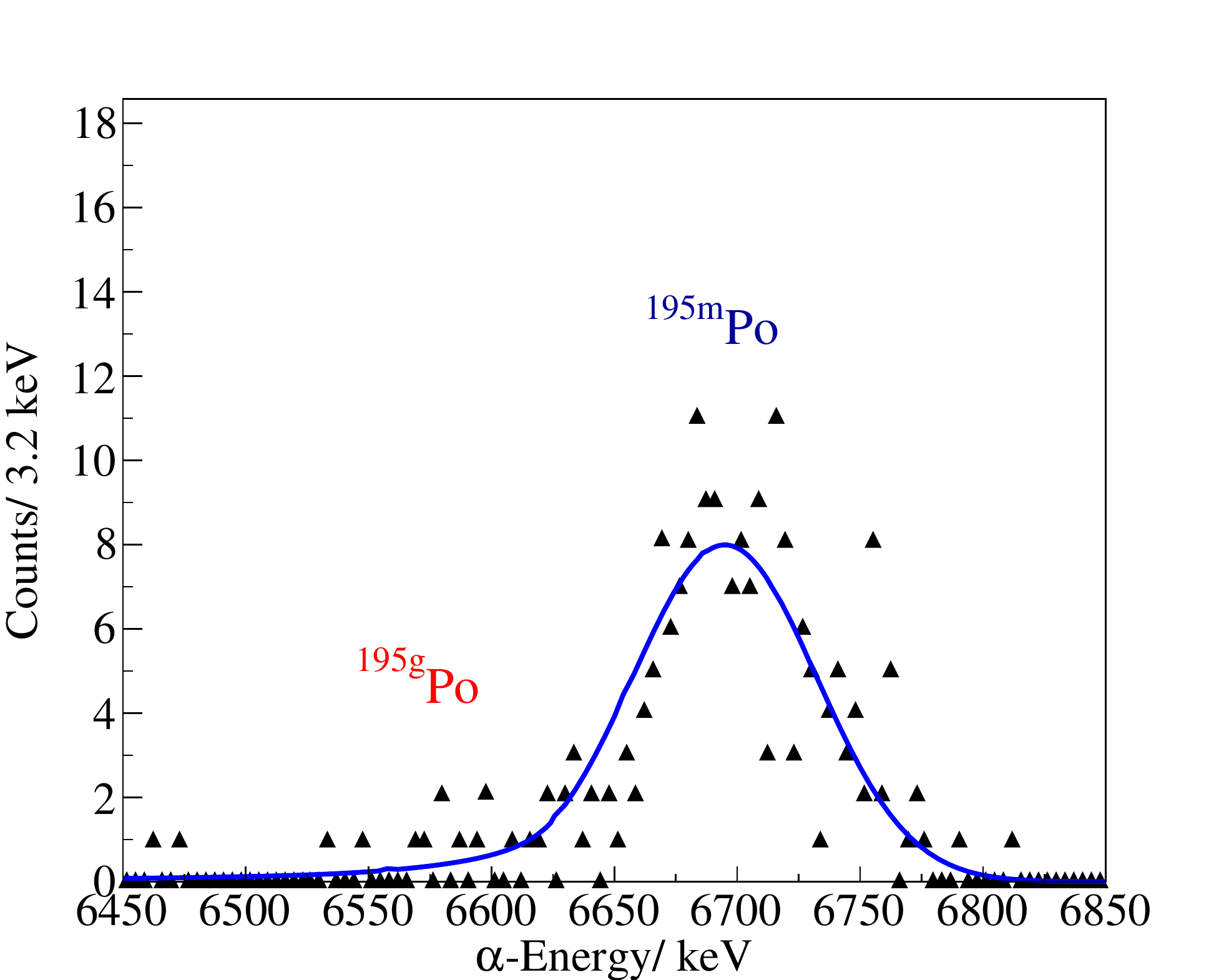}}
\subfloat[ ]{\includegraphics[width = 0.45\textwidth]{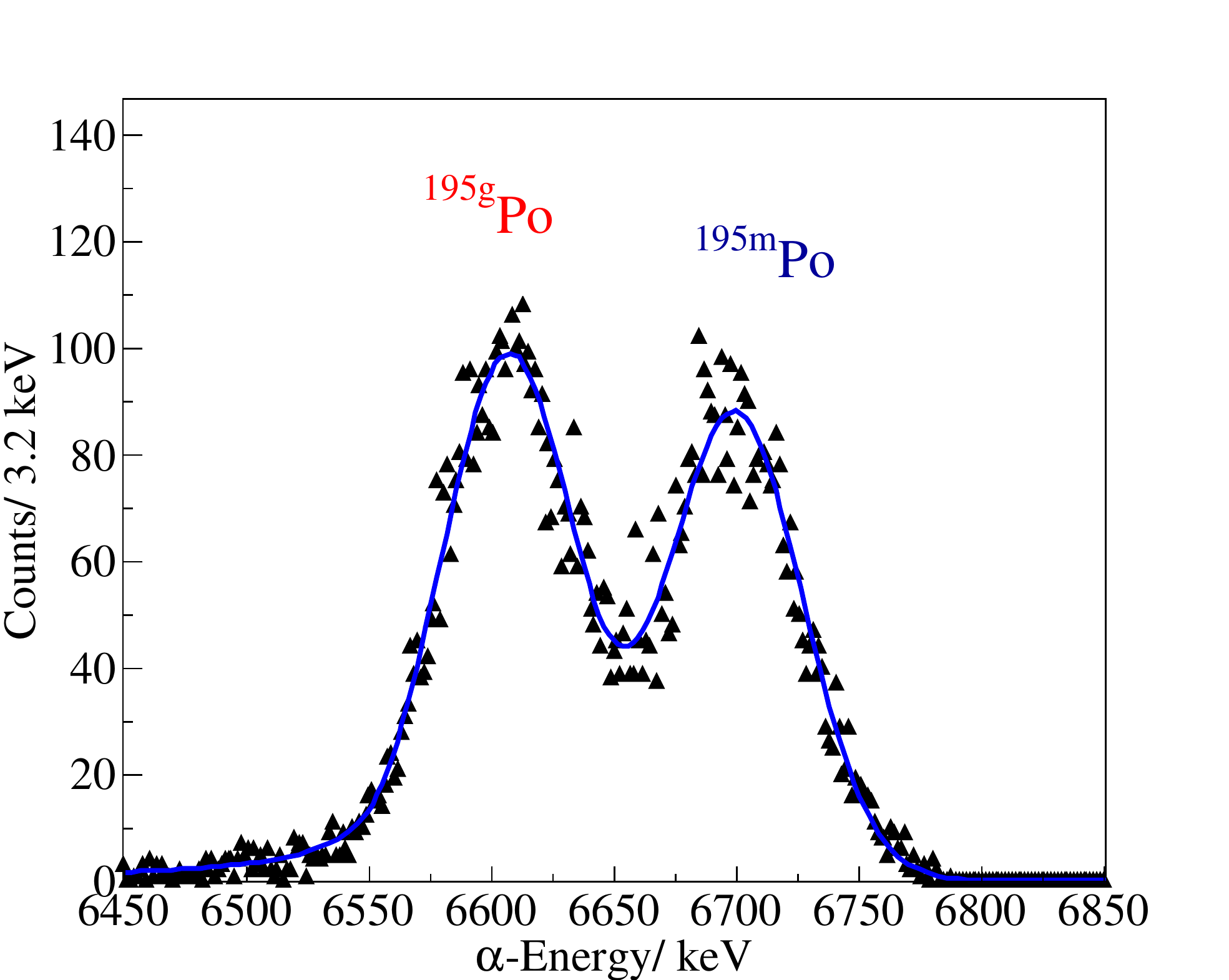}} \\
\subfloat[ ]{\includegraphics[width = 0.45\textwidth]{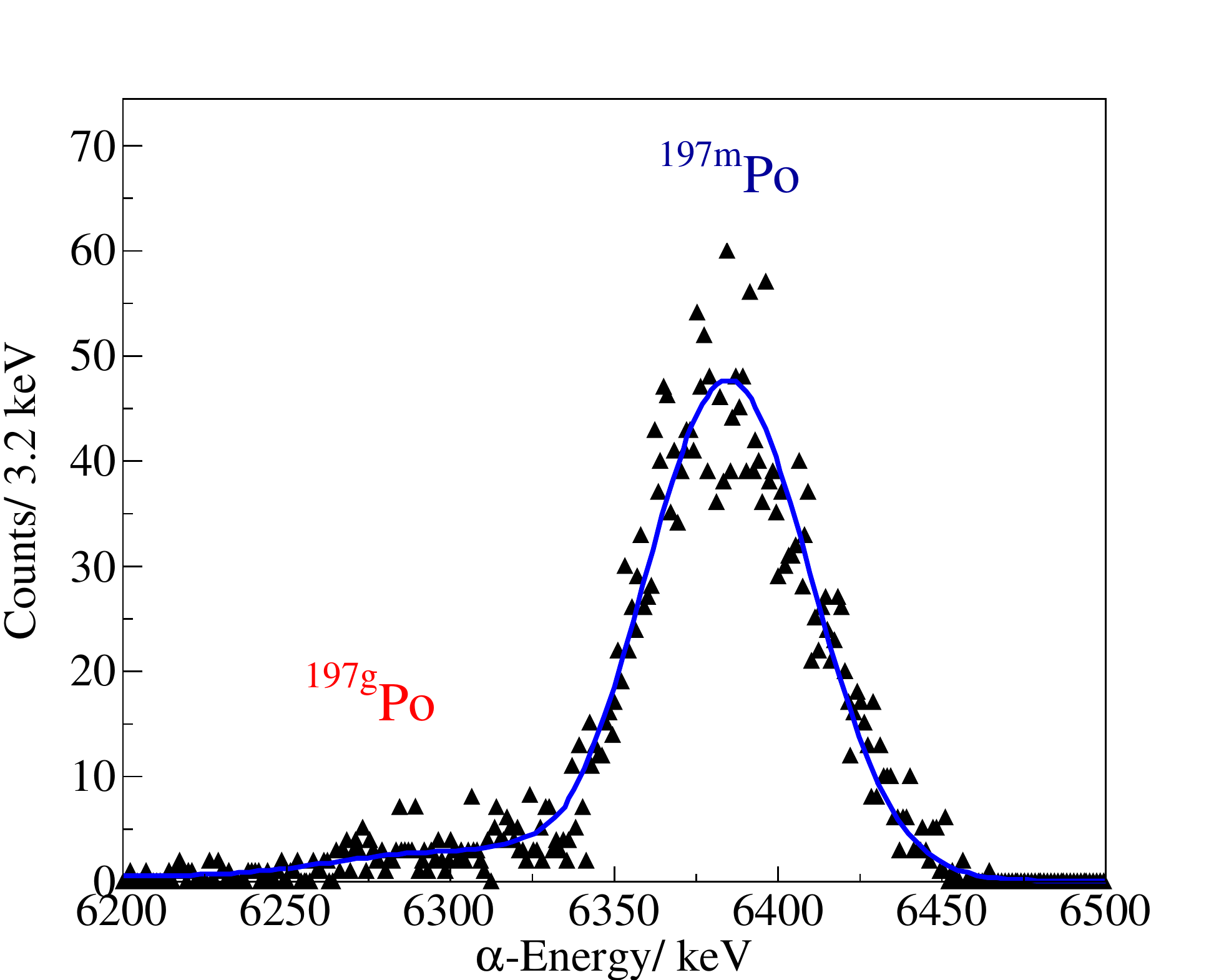}}
\subfloat[ ]{\includegraphics[width = 0.45\textwidth]{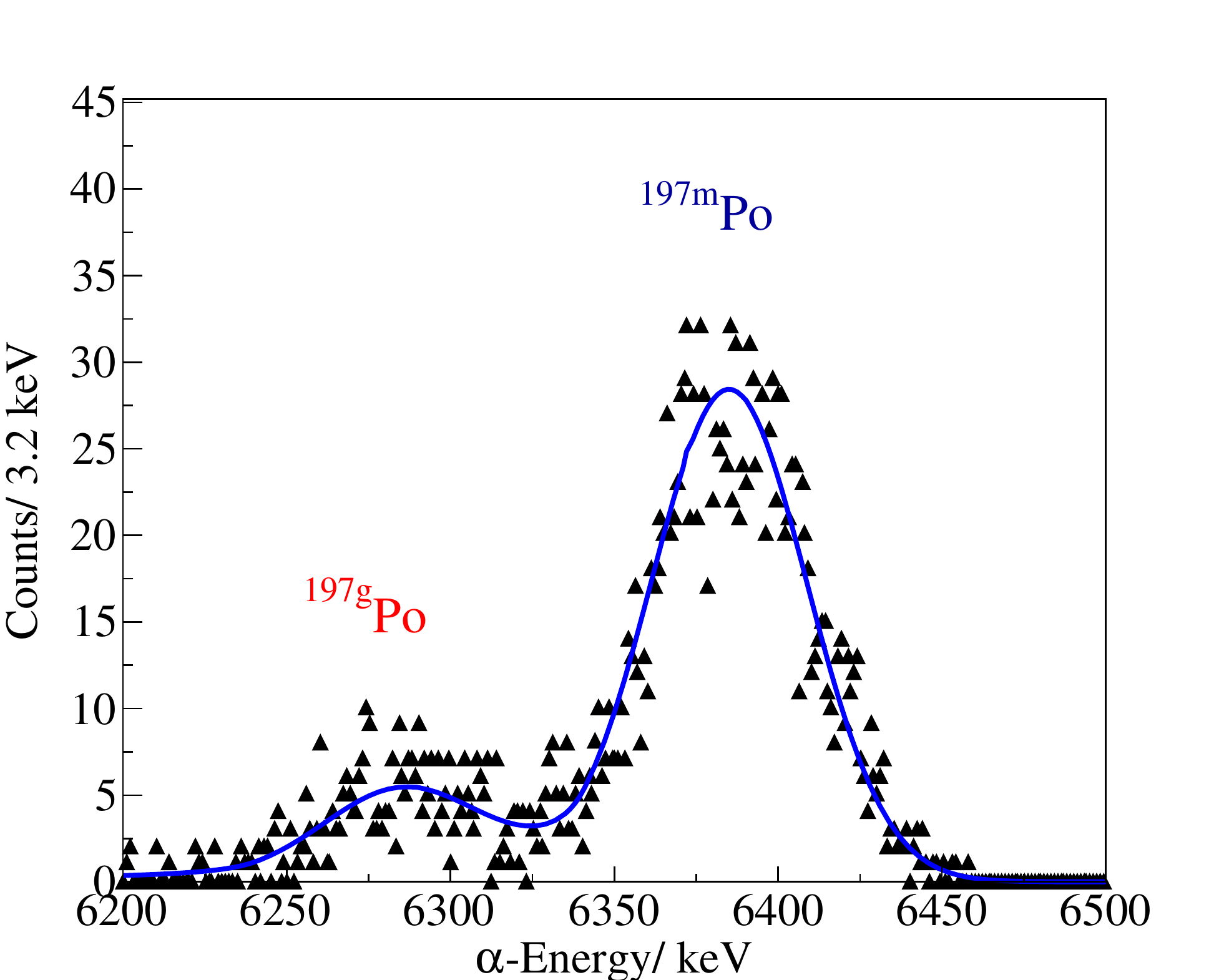}}
\end{tabular}
\caption{ (color online) The $\alpha$-decay energy spectra for $^{195}$Po, (a) and (b), and $^{197}$Po, (c) and (d), measured with the two silicon detectors surrounding the implantation point in DSS2.0. The spectra were recorded at different RILIS wavenumbers: (a) and (c) taken at 11853.75 cm$^{-1}$ and 11853.95 cm$^{-1}$ respectively, while (b) and (d) at 11853.83 cm$^{-1}$.}
\label{fig:3}
\end{figure*}

\begin{figure}[t!]
\includegraphics[width=0.7\textwidth]{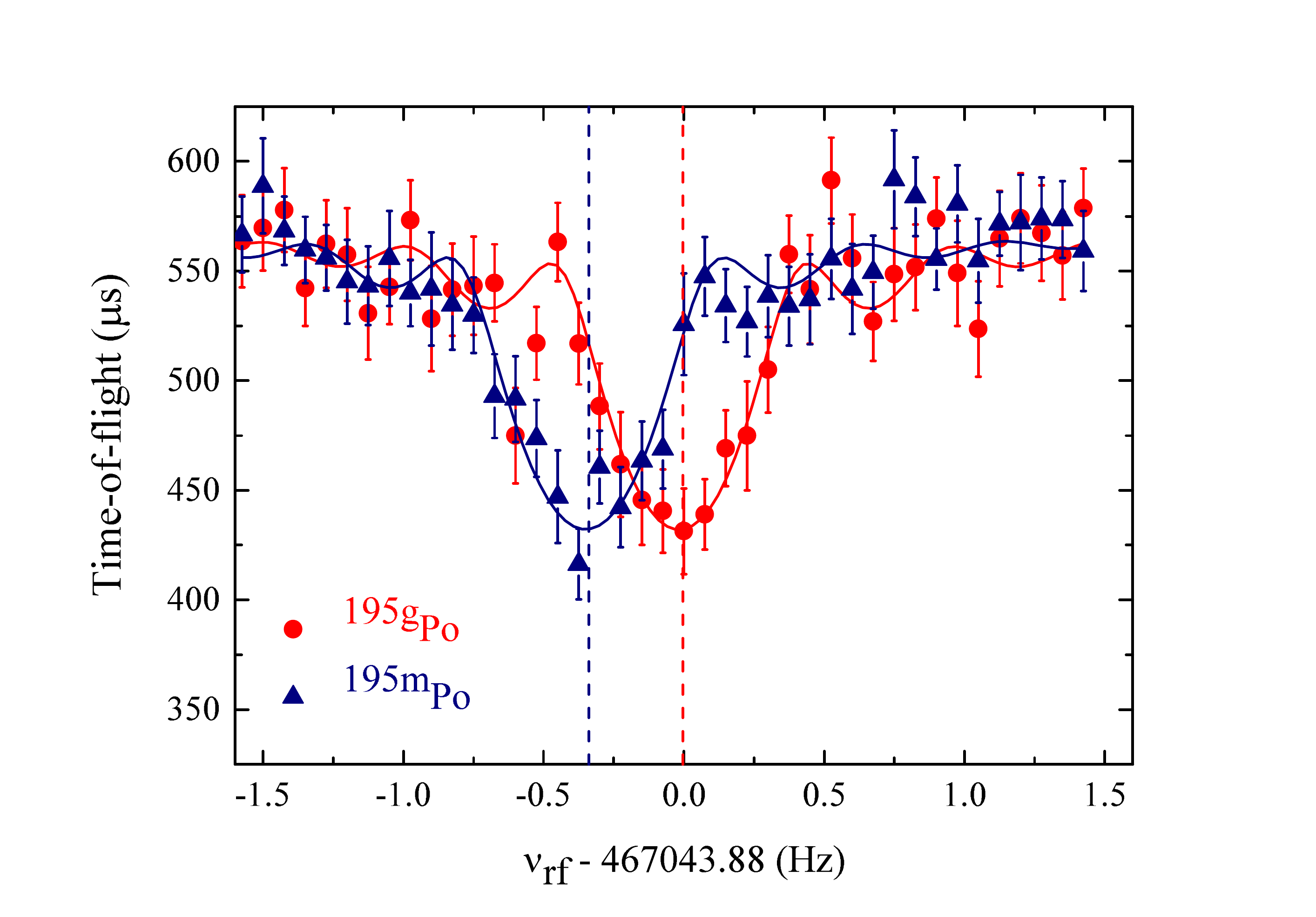}
\caption{\label{fig:4} (color online) ToF-ICR resonances of $^{195(g,m)}$Po
 with 2\,s excitation time in blue triangles and red circles, respectively. The red (blue) dashed line represents the center frequency of the ground (excited) state. Quasi-pure samples of either state were measured thanks to selective enhancement from RILIS and different half-lives (see text for details). The solid lines represent the fit of the theoretical line shape \cite{18} to the experimental data.}
\end{figure}

For the {$^{195}$Po} measurements, it was  possible to selectively enhance  the production of either nuclear state by tuning the laser to a specific wavenumber. In case of the high-spin state $13/2^{(+)}$ with half-life of 2\,s, the state selection was achieved by tuning the laser frequency to the wavenumber 11853.75 cm$^{-1}$ where the ionization of the $13/2^{(+)}$ state is dominant, see Fig.~\ref{fig:2}(a). The amount of the $13/2^{(+)}$ state was deduced from the recorded $\alpha$-decay energy spectrum to be $R_{m}=94(2)\%$  as illustrated in Fig.~\ref{fig:3}(a). Thus, three quasi-pure high-spin ToF-ICR resonances were recorded with excitation times of up to 2\,s. A representative ToF-ICR spectrum is shown in Fig.~\ref{fig:4}. 

For the longer-lived low-spin state $3/2^{(-)}$ with half-life of 4.64\,s, the highest intensity corresponds to the wavenumber 11853.83 cm$^{-1}$. This laser setting is not sufficient to separate the two states but rather gives a mixture between the two states leading to $R_{g}=56(1)\%$ ground state, as inferred from the $\alpha$-decay energy spectrum in Fig.~\ref{fig:3}(b). Therefore, the difference in the half-life of the two states of $^{195}$Po was exploited, to modify the ratio between them.  In addition to the RILIS laser setting, the ions  were kept in the preparation trap for more than 2\,s and then transferred to the precision Penning trap. This way, $R_{g}$ was increased and three ToF-ICR resonances were recorded with excitation times of up to 2\,s in the precision Penning trap, shown in Fig.~\ref{fig:4}. For each state, the frequency ratio is presented in Table \ref{ta:1}.

\begin{figure}[h!]
\includegraphics[width=0.7\textwidth]{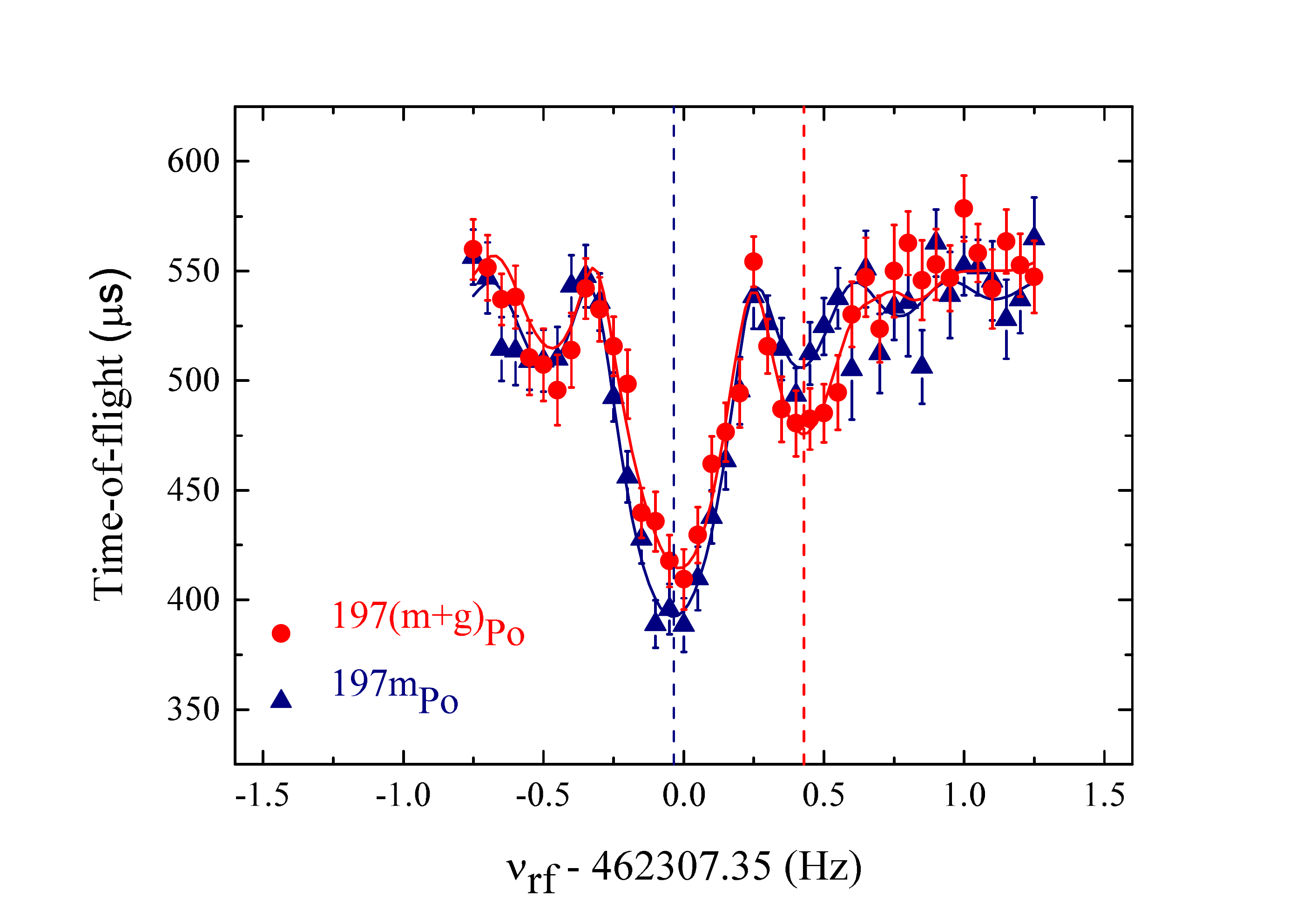}
\caption{\label{fig:6} (color online) ToF-ICR resonances of $^{197(g,m)}$Po with 5\,s excitation time. The two resolved resonances in the red circle ToF-ICR spectrum are isomer (left) and ground (right) states. The pure isomeric state is represented by the blue triangles ToF-ICR  spectrum. The red (blue) dashed line is the center frequency of the ground (excited) state. The solid lines are fits of the experimental data to the theoretical line shape \cite{18}. }
\end{figure}

\subsection{$^{197}$Po}
In the case of the longer-lived $^{197}$Po with half-life of 84\,s, it was possible with the RILIS laser tuned to a wavenumber of 11853.95 cm$^{-1}$ (see Fig.~\ref{fig:2}(b)) to obtain pure high-spin ToF-ICR resonances with an excitation time of 3\,s. At this laser setting, no evidence for the low-spin in the $\alpha$-decay energy spectrum is found (see Fig.~\ref{fig:3}(c)). This is due to the use of the ionization scheme based on the Rydberg  state, which reduced the power broadening of the lines. This resulted in less contamination from the ground state than the previous study. On the contrary, the hyperfine structure of the low-spin state overlaps with the high-spin state as shown in Fig.~\ref{fig:2}(b) and is present in the $\alpha$-decay energy spectrum of Fig.~\ref{fig:3}(d). By using the wavenumber 11853.83\,cm$^{-1}$, the beam content for the $3/2^{(-)}$ state was $R_{g}=25(1)\%$ as obtained from the DSS2.0. Fully-resolved resonances  of the isomer and ground states were achieved with a quadrupole excitation time of 5\,s  as shown in Fig.~\ref{fig:6}. In order to correct the shift in the cyclotron frequency center due to simultaneously trapped isomeric and ground state ions, the maximum number of detected ions was limited to 2 in each measurement cycle. For each state, the frequency ratio $ \bar{r}$ is presented in Table \ref{ta:1}.

\begin{figure}[h!]
\includegraphics[width=0.7\textwidth]{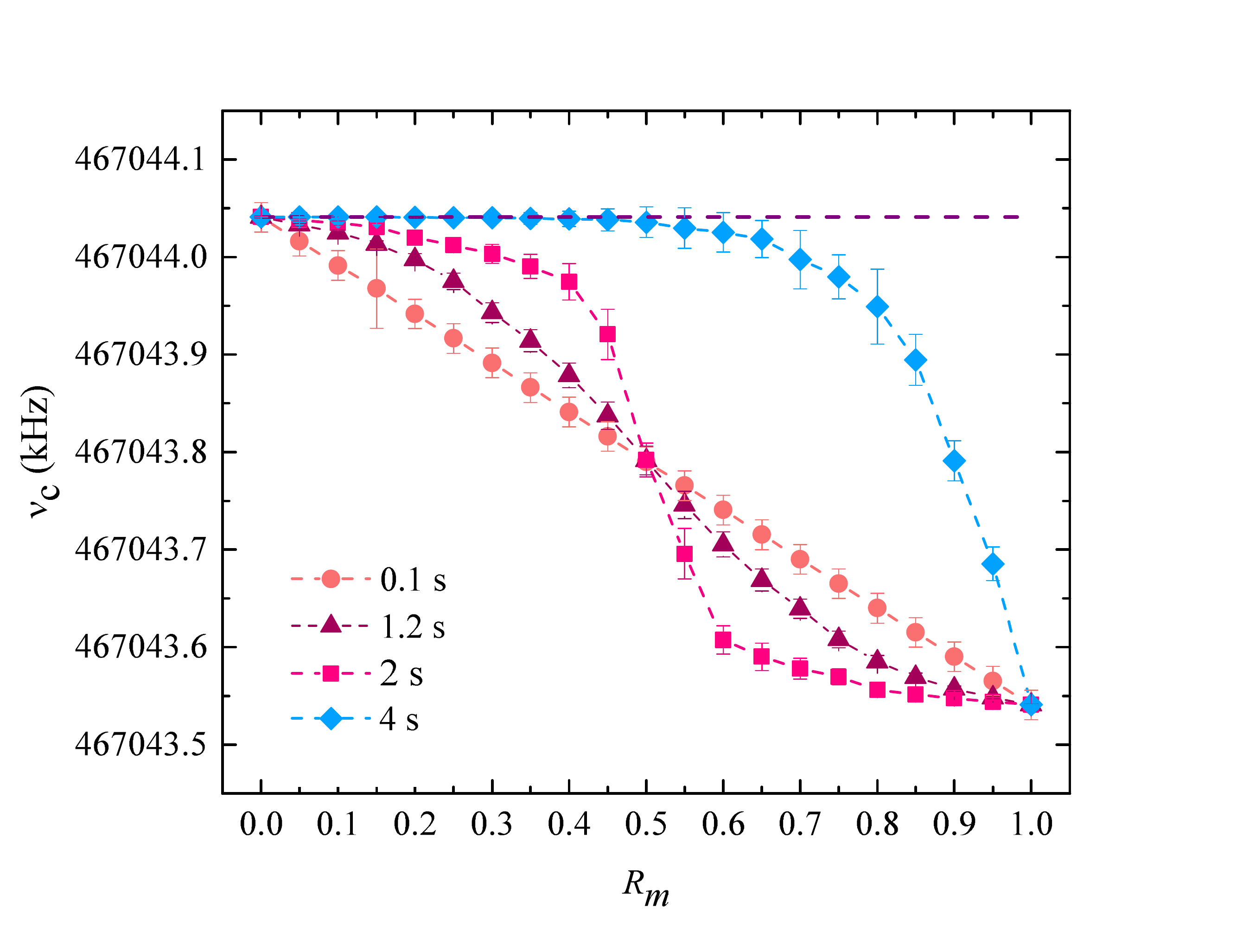}
\caption{\label{fig:11} (color online) Center frequency $\nu_{c}$ resulting from a fit of a mixed resonance with a single-resonance curve, as a function of the isomeric state ratio $R_{m}$. It was obtained from simulated data for different excitation times. The purple dashed straight line represents the pure cyclotron frequency of the ground state where the ratio of ground-to-isomeric state is 1:0. When two resonances are clearly identified (for longer excitation times), the resonance of interest is systematically selected. See text for details.}
\end{figure}
\begin{table*}[t!]
\begin{ruledtabular}
\def\arraystretch{1.5}
\caption{Weighted-average frequency ratios $ \bar{r}$ of Po$^{+}$ isotopes obtained
with respect to $^{133}$Cs$^+$ \cite{Audi2012,Wang2012}. The corresponding mass excess values (ME) are shown as well. AME12 values were taken from \cite{Audi2012,Wang2012}.}\label{ta:1}
\begin{tabular}{ccccc}
\toprule
Isotopes& $I^\pi$& $ \bar{r}$&~~ ME ($\mathrm{keV}$)&~~AME12 ($\mathrm{keV}$)\\ 
\hline
$^{195g}\mathrm{Po}$&3/2$^{(-)}$& $1.4671205566(495)$&~ $-11118(6)$&~~$-11060(40)$ \\
$^{195m}\mathrm{Po}$& 13/2$^{(+)}$& $1.4671217650(584)$&~~ $-10968(7)$&~~ $-10960(64)$ \\ 
$^{196g}\mathrm{Po}$& 0$^+$& $1.4746257515(477)$&~~ $-13468(6)$&~~ $-13483(13)$ \\
$^{197g}\mathrm{Po}$& (3/2$^-$)& $1.4821505060(804)$&~~ $-13396(10)$&~~ $-13360(50)$ \\
$^{197m}\mathrm{Po}$& 13/2$^{(+)}$& $1.4821521160(523)$&~~$-13197(7)$&~~$-13130(94)$\\  
$^{199m}\mathrm{Po}$& (13/2$^+$)& $1.4971864768(395)$&~~ $-14929(5)$&~~ $-14904(23)$ \\
$^{203g}\mathrm{Po}$& 5/2$^-$& $1.5272639530(453)$&~~ $-17310(6)$ &~~$-17311(9)$\\
$^{208g}\mathrm{Po}$& 0$^+$& $1.5648835996(448)$&~~$-17463(6)$&~~ $-17469(2)$\\ \bottomrule
\end{tabular}
\end{ruledtabular}
\end{table*}
\subsection{Analysing mixed resonances}
The relative uncertainty associated with the measurements is evaluated as a result from the drift of the magnetic field the trap is placed in, a mass-dependent effect, and a residual systematic uncertainty \cite{A. Kellerbauer2003}. Other possible sources of systematic errors have also been considered, resulting from the existence of a small fraction of partially unresolved ground/isomeric state in the $^{195g,m,197m}$Po measurements, respectively, which were analyzed as single resonances. To this end, for each data file in question we generated an idealized pure resonance with parameters (magnetron radius, conversion) fixed from the experimental spectrum. A second resonance was then generated, having the same parameters but being a mixture of the ground and isomeric resonances with the weight given by the $\alpha$-spectroscopy information. The frequencies used for the pure and mixed resonances were the ones obtained from the analysis of the data files using single fits. The mixed resonance was then also evaluated using a single fit and the shift between the resulting center frequency and the center of the pure resonance was added in quadrature as a systematic error. 

The more detailed evaluation of systematic errors resulting from mixed resonances fitted as single resonances is shown in Fig.~\ref{fig:11}. A systematic shift of the fitted ground-state frequencies was observed as the weight of the isomeric state varies from 0 (no isomeric state is present) to 1 (purely isomeric state). When the two resonances are fully resolved, e.g. with a longer excitation time than 4\,s, no change in the frequency is seen as $R_{m}$ increases. The change becomes nonlinear when the excitation time is reduced to 2\,s. The results also show that a shorter excitation time leads to linear changes since the two resonances are merged. 

The frequency ratio in these measurements were determined with respect to $^{133 }$Cs$^+$. The weighted-average of the frequency ratios and the corresponding mass excess (ME) values obtained from these measurements are presented in Table \ref{ta:1}. The low-spin ME values of $^{195,197}$Po agree with AME2012 \cite{Audi2012,Wang2012} within one standard deviation and provide more precise values.

 \subsection{$^{199m}$Po}
 The measurements of $^{199}$Po isotopes were performed using RILIS in broadband mode to produce a balanced ratio between the (13/2$^+$) and the (3/2$^-$) states. The $\alpha$-decay measurements at this setting gives $R_{m}=96(1)\%$, in favor of the high-spin (13/2$^+$). Thus, it was not possible to obtain a favorable ratio for the (3/2$^-$) state. However using a laser wavenumber of 11854.02 cm$^{-1}$ in narrowband mode, the beam quality increased to $R_{m}=99.1(3)\%$, with high purity for the (13/2$^{+}$) state. The mass of this state was obtained by four ToF-ICR measurements with an excitation time of 3\,s, one of which was recorded with the broadband laser configuration. The resulting weighted-average frequency ratio and ME are displayed in Table \ref{ta:1}.
 
 \subsection{$^{196,203,208}$Po}
 In addition to $^{195,197,199}$Po, the masses of $^{196,203,208}$Po were determined using the precision Penning trap. Two resonances were recorded with the ToF-ICR method for $^{196}$Po  and three for $^{203,208}$Po. The values of $^{196,203,208}$Po agree with the AME2012 values \cite{Audi2012,Wang2012}  within one standard deviation. The resulting weighted-average frequency ratios and the ME are displayed in Table \ref{ta:1}.
 
 \begin{table}[h]
 \begin{ruledtabular}
\def\arraystretch{1.5}
\caption{Excitation energies of the isomers in odd-$A$ polonium isotopes determined
from mass measurements of the $3/2^{(-)}$ and 13/2$^{(+)}$ states. The excitation energies of other isotopes linked by $\alpha$-decay chains are also presented.}  
\label{ta:2}
\begin{center}
\begin{tabular}{*{3}{c}}

 Isotopes&~~~~~ $I^\pi$&~~~~~~ $E^*~(\mathrm{keV})$\\
\hline
$^{205}\mathrm{Ra}$&~~~~~~13/2$^{(+)}$& ~~~~~~$  278(31)  $ \\
$^{203}\mathrm{Ra}$&~~~~~~13/2$^{(+)}$& ~~~~~~ $ 246(17)   $ \\  
$^{201}\mathrm{Rn}$& ~~~~~~13/2$^{(+)}$&~~~~~~ $  248(12)  $ \\
$^{199}\mathrm{Rn}$& ~~~~~~13/2$^{(+)}$&~~~~~~ $  222(13)  $ \\  
$^{197}\mathrm{Po}$& ~~~~~~13/2$^{(+)}$&~~~~~~ $  199(11)  $ \\
$^{195}\mathrm{Po}$& ~~~~~~13/2$^{(+)}$&~~~~~~$   150(10) $ \\  
$^{193}\mathrm{Pb}$& ~~~~~~13/2$^{(+)}$&~~~~~~$  95(28)  $ \\
$^{191}\mathrm{Pb}$& ~~~~~~13/2$^{(+)}$&~~~~~~ $  55(12)  $ \\ 
\end{tabular}
\end{center}
\end{ruledtabular}
\end{table}

\section{\label{Disc}Discussion}

The $\alpha$-decay spectroscopy supported the mass measurements by providing an accurate measurement of the high-to-low-spin ratio for a given RILIS laser frequency setting. The ratio was used to identify the spin of the nuclear state of the ions for the corresponding ToF-spectrum.
Furthermore, it was possible to distinguish the state ordering in the mass measurement. As follows from Eq.~\eqref{M0}, the state with a higher cyclotron frequency corresponds to the lighter, more bound, ground state, while the one with a lower cyclotron frequency corresponds to the excited state with higher mass. The results obtained from $\alpha$-decay spectroscopy and mass measurements combined show that the $13/2^{(+)}$ state is the excited state in both $^{195}$Po and $^{197}$Po.

The excitation energy $E^*$ of the  $13/2^{(+)}$ isomer state in $^{195,197}$Po was obtained in terms of the measured cyclotron frequency using Eq.~\eqref{M1}.
 \begin{equation}
\label{M3}
E^*= (\bar{r}_{iso}-\bar{r}_g) (m_{ref}-m_e),
\end{equation}
where $ \bar{r}_{iso}$ and $\bar{r}_g$ are the weighted-average of the frequency ratios of isomeric and ground states, respectively.

\begin{figure}[t!]
\centering
\includegraphics[width=0.7\textwidth]{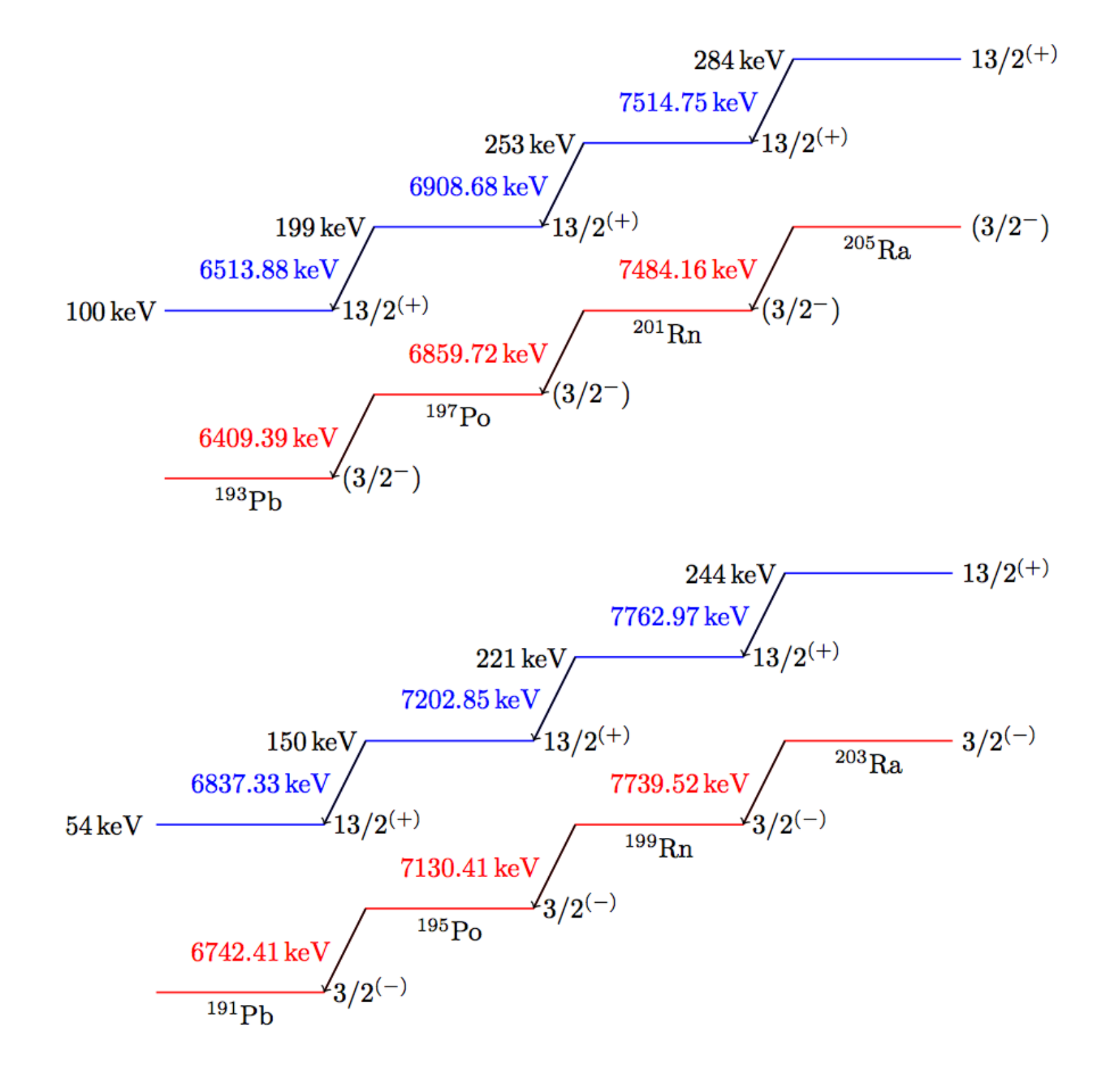}
\caption{(color online) Simplified $\alpha$-decay scheme from low- and high-spin states in  $^{203,205}$Ra to $^{191,193}$Pb. Using the excitation energies from this work in $^{195,197}$Po and the $Q_\alpha$-values
\cite{Audi2012,Wang2012}, the excitation energy in each linked isotope is determined. The red and blue lines represent the ground and excited states, respectively.}
\label{fig:12}
\end{figure}
\begin{figure}[t!]
\centering
\includegraphics[width=0.7\textwidth]{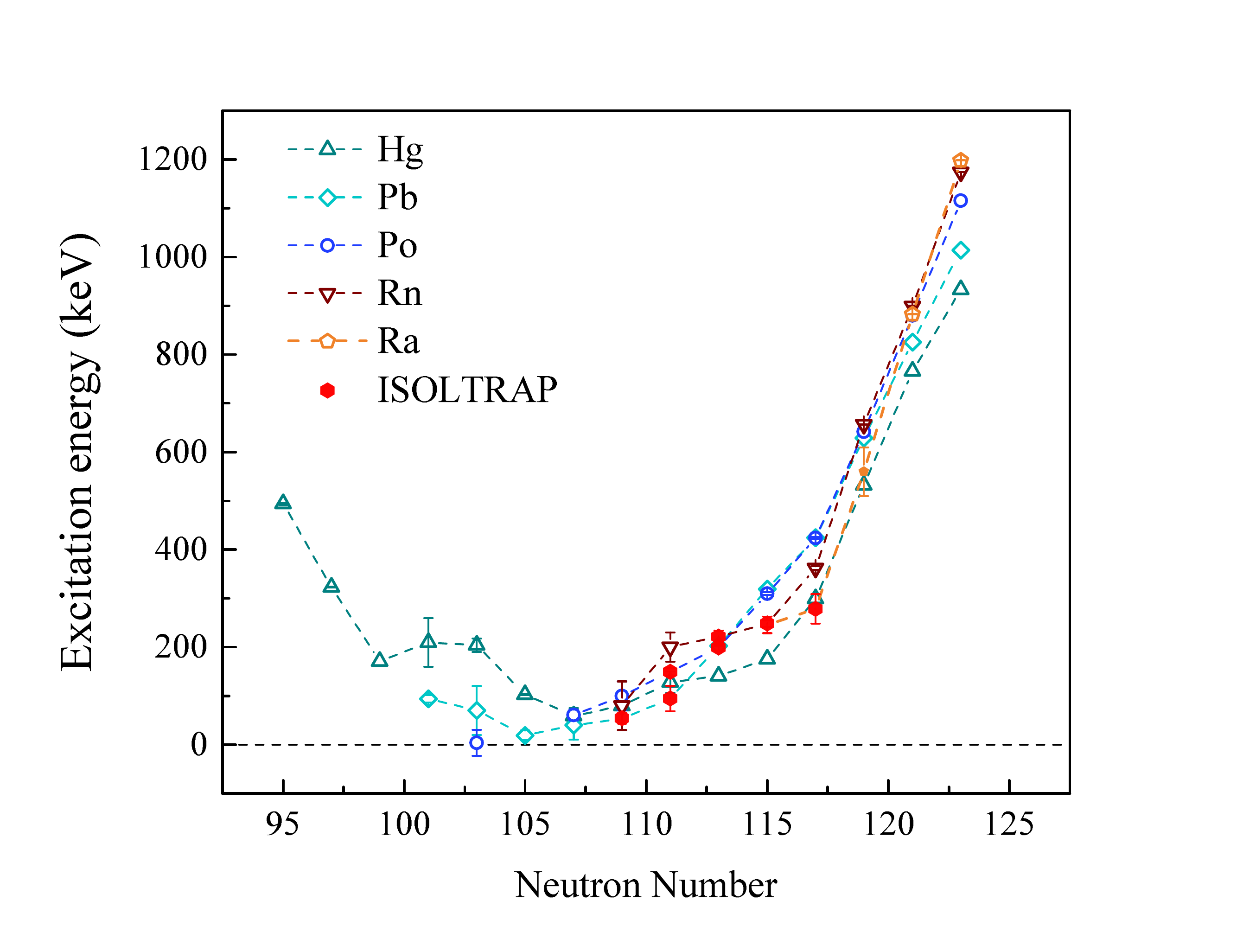}
\caption{\label{fig:8} (color online) The excitation energy of the $13/2^{(+)}$ state in odd-$A$  polonium isotopes in the neutron-deficient lead region. The excitation energies determined indirectly by using the new mass data and  $\alpha$-decay $Q_\alpha$-values are also shown. The results obtained from ISOLTRAP (red) show that the $13/2^{(+)}$ configuration does not become the ground state. Data from \cite{Audi2012,Wang2012} are shown with other colors.}
\end{figure}

Based on the extracted excitation energy of the $13/2^{(+)}$ state in $^{195,197}$Po, the excitation energies of other isotopes in the $\alpha$-decay chains shown in Fig.~\ref{fig:12} are determined by adding or subtracting the difference in $Q_\alpha$-value between the two high-spin and low-spin chains. According to the new ISOLTRAP results, the high-spin state never becomes the ground state. The $13/2^{(+)}$ state is the excited state in each of $^{191,193}$Pb, $^{199,201}$Rn and $^{203,205}$Ra. All $13/2^{(+)}$ excitation energies are presented in Table \ref{ta:2}. Figure \ref{fig:8} shows the $13/2^{(+)}$ excitation energy across a long chain of isotopes in the lead region. The isomers are obtained by a valence neutron occupying the $\nu i_{13/2}$ orbital, as evidenced in Hg, Pb, and Po by the nuclear dipole magnetic moment \cite{Hg,Seliverstov2009,Seliverstov2014}. The evolution of their energy is therefore expected to relate to that of the single particle orbital $\nu i_{13/2}$. A parabolic trend is observed, reaching a minimum at $N=107$ for mercury, $N=105$ for lead and $N=103$ for polonium.

This behaviour is also observed in the neutron-rich isotopes near the tin chain with (proton number $Z = 50$).  The $\nu h_{11/2}$ state is observed as an isomer in neutron-rich tin ($Z=50$), cadmium ($Z=48$) and tellurium ($Z=52$) isotopes. The energy of the excited state arising from the $\nu h_{11/2}$ orbital drops from $N=82$ and almost becomes degenerate with the ground states $\nu s_{1/2}$ and $\nu d_{3/2}$ for nuclei between $N=65$ and $82$.  In the tin isotopes, however, the difference in energy between the $11/2^{-}$ and $3/2^{+}$ states decreases with smaller $N$, which leads to level inversion as the isomeric state $11/2^{-}$ becomes the ground state between $N=73$ and $N=77$.

It is also interesting to note that the energies of the $13/2^+$ states are generally independent of the occupation of the proton orbitals. They all behave alike whether for a closed shell (Pb), two proton holes below $Z=82$ (Hg), or have protons in orbitals above $Z=82$ (Po, Rn, and Ra). It is even more striking as this is not seen in other observables, such as the electromagnetic moments where each isotopic chain behaves differently \cite{Seliverstov2009,Seliverstov2014}. The particular evolution of the energy of the 13/2$^+$ state, relative to the ground 3/2$^-$ state, exhibits a similar drop in energy to the first excited $0^+$ state in the even-even nuclei as a function of decreasing neutron number. From a spherical core close to $N=126$, the incoming oblate energy minimum drops down to $N\sim107$, and then exhibits a rather flat behaviour moving to still lower neutron numbers, even when crossing the mid-shell point. This close correlation indicates a strong interplay between the odd-particle moving in the average field of a changing even-even core. The deformation values extracted from experimental data (such as the magnetic dipole and electric quadrupole moments), when available, are consistent with the above results \cite{Seliverstov2009,Seliverstov2014,Ulm1986}. The charge radii and root mean-square values for the quadrupole deformation are rather well reproduced by mean-field studies \cite{DeWitte2007,Cocolios2011,Kesteloot2015}.

\section{\label{Conc}Conclusion}
Direct mass measurements of the low and high-spin states in the neutron-deficient isotopes $^{195}$Po and $^{197}$Po have been performed. The hyperfine structure of the radioactive $^{195,197}$Po and the difference in the half-lives between the ground and isomeric states of $^{195}$Po,  also allowed mass measurements for three of the states. With purified ion ensembles the mass values of $^{(195,197)m,g}$Po were determined by Penning-trap mass spectrometry, from which we obtained for the first time the state ordering and the excitation energy of the $13/2^{(+)}$ state in $^{195,197}$Po. The connection between the measured mass values and the nuclear states was performed by adding an $\alpha$-decay-spectroscopy setup behind the MR-ToF MS, profiting from its resolving power to obtain pure polonium samples for increased sensitivity. This is the first application of MR-ToF-assisted decay-spectroscopy at ISOLTRAP.

Additionally, the $13/2^{(+)}$ excitation energy of the polonium isotopes' $\alpha$-decaying daughters and parents was determined. 
Our presented mass measurements provide information on the behaviour of the $13/2^{(+)}$ states in the lead region near $N=104$, which remain an excited state as driven by the pairing force in high-$j$ orbitals. This is in contrast to the $11/2^{-}$ state in the Sn region which becomes the ground state between $N=73$ and $N=77$.

\section{Acknowledgements}
We would like to acknowledge the ISOLDE Collaboration and technical teams. We would like to thank the CRIS Collaboration for the use of the DSS2.0 and the GSI target laboratory for providing the carbon foils. 

This work was supported by 
the German Federal Ministry for Education and Research (BMBF) (Grant Nos. 05P09ODCIA, 05P15HGCIA, 05P12HGCI1, 05P12HGFNE, and 06MZ215), 
a Marie Curie Actions grant (No. MEIF-CT-2006-042114, EU FP6 program), 
the Max-Planck Society, 
the Helmholtz Alliance, 
the European Union seventh framework through ENSAR (Contract No. 262010), 
the French IN2P3, 
the Spanish MICINN (Grant Nos. FPA-2010-17142 and FPA-2011-24553), 
the UK Science and Technology Facilities Council (STFC), 
the Robert-Bosch-Foundation, 
the IUAP-Belgian State Belgian Science Policy (BRIX network P7/12), 
FWO-Vlaanderen (Belgium), 
and GOA 10/010 from KU Leuven. N.A.A was supported by a scholarship from Al-jouf University.

\end{document}